\documentclass{article}
\usepackage{amsmath,epsfig,amsfonts}
\usepackage[preprint]{spconfa4}
\usepackage{algorithm,algorithmicx,algpseudocode}

\usepackage{rotating}
\usepackage{caption}
\usepackage{multirow}
\usepackage{multicol}
\usepackage{cuted}
\usepackage{bbding}

\copyrightnotice{\textbf{978-1-7281-1331-9/20/\$31.00~\copyright 2020 IEEE}}

\let\OLDthebibliography\thebibliography
\renewcommand\thebibliography[1]{
  \OLDthebibliography{#1}
  \setlength{\parskip}{0pt}
  \setlength{\itemsep}{0pt plus 0.3ex}
}

\newcommand{\ShrinkedSection}[1]{
	\vspace{-.3cm}
	\section{#1}
	\vspace{-.2cm}
}
\newcommand{\ShrinkedSubsection}[1]{
	\vspace{-.1cm}
	\subsection{#1}
	\vspace{-.15cm}
}

\begin{document}\sloppy

	\def\x{{\mathbf x}}
	\def\L{{\cal L}}

	\title{Statistical Detection of Collective Data Fraud}
	%
	\name{Ruoyu Wang$^{1,2}$, Xiaobo Hu$^{3}$, Daniel Sun$^{2,4,5}$\Envelope, Guoqiang Li$^{1,4}$\Envelope, Raymond Wong$^{2}$, Shiping Chen$^{5}$, Jianquan Liu$^{6}$}
	\address{$^{1}$Shanghai Jiao Tong University, Shanghai 200240, China, \{wang.ruoyu, li.g\}@sjtu.edu.cn;\\
		$^{2}$University of New South Wales, Sydney, Australia, wong@cse.unsw.edu.au;\\
		$^{3}$Peking University, Beijing, China, 1901210305@pku.edu.cn;\\
		$^{4}$Enhitech Co. Ltd., Shanghai 200241, China;\\
		$^{5}$Data61, CSIRO, Australia, \{daniel.sun, Shiping.Chen\}@data61.csiro.au;\\
		$^{6}$Biometrics Research Laboratories, NEC Corporation, Japan, jqliu@nec.com.}

	\maketitle
	\pagenumbering{gobble}

	\begin{abstract}
		Statistical divergence is widely applied in multimedia processing, basically due to regularity and interpretable features displayed in data. However, in a broader range of data realm, these advantages may no longer be feasible, and therefore a more general approach is required. In data detection, statistical divergence can be used as a similarity measurement based on collective features.
		In this paper, we present a collective detection technique based on statistical divergence.
		The technique extracts distribution similarities among data collections, and then uses the statistical divergence to detect collective anomalies.
		Evaluation shows that it is applicable in the real world. 
	\end{abstract}

	
	\ShrinkedSection{Introduction}
		\label{sec:intro}
		Statistical divergence is widely applied in multimedia processing. Prevalent applications include multimedia event detection~\cite{amid2014unsupervised}, content classification~\cite{moreno2004kullback,park2005classification} and qualification~\cite{pheng2016kullback,goldberger2003efficient}. It has been attracting more attention since the dawn of big data era, basically due to regularity and interpretable features displayed in the data. However, in a broader range of data realm, these advantages may no longer be feasible (e.g. in online sales data records). It requires a more general approach.
		
		General data frauds can be caused by manipulation from outside hackers.
		\textit{Data Manipulation} here, according to a NSA definition, refers to behaviours which ``change information contained in those systems, rather than stealing data and holding it for ransom''.
		In 2013, hackers from Syria put up fake reports via Associated Press' Twitter account and caused a 150-point drop in the Dow~\cite{SyriaHacker}.
		
		It is hard to detect a single record that is altered but still remains in correct value scopes, but if sufficient data records are altered to change a final decision, we can still detect malicious data manipulation behaviours.
		According to our observation, typical manipulations on numerical data will lead to a drift or distortion of its original distribution.
		To address problems caused by data manipulation, we proposed a novel technique which sorts out manipulated data collections from normal ones by adopting statistical divergence.
		In this paper, we focus on a concrete data manipulation problem: click farming in online shops, and try to apply our technique to pick out dishonest behaviours.
		Our technique maps data collections to points in distribution spaces and reduce the problem to classical point anomaly detection.
		Optimizations estimate ground truth, mapping each data collection into a single real number within a definite interval. Then a Gaussian classifier can be applied to detect outliers derived from manipulated data. To automatically calculate adaptive threshold for the classifier, we keep two evidence sets for both normal points and anomalies, taking advantage of the property provided by statistical divergence.
		In the dynamic environments, these evidence sets act intuitively as slide windows and keep up to the evolving features in dynamic scenarios.
		Our contribution includes: 1) A brief review on data fraud detection and a study on the problem of click farming; 2) Detailed description of both basic and optimized framework of our technique, resolving several technical difficulties such as automated adaptive threshold; 3) Comprehensive experiments that test efficiency of our technique and a comparison with previous work on similar topic.
		
		The rest of the paper is organised as follows: Section~\ref{sec:related-work} states related work on data anomaly detection and briefly introduces click farming.
		Details of proposed technique are introduced in section~\ref{sec:algorithm-details}. Then section~\ref{sec:evaluation} presents evaluation results and further findings of the algorithm. Finally, the paper is concluded in section~\ref{sec:conclusion}.
	
	\ShrinkedSection{Related Work}\label{sec:related-work}
		\ShrinkedSubsection{Data Anomaly Detection}
			Statistical divergence was applied mainly as classifiers on multimedia content~\cite{park2005classification}, especially as kernels in SVMs~\cite{moreno2004kullback}. As a similarity measurement, it can also be used in qualitative and quantitative analysis in image evaluation~\cite{pheng2016kullback,goldberger2003efficient}. \cite{amid2014unsupervised} adopted divergence to detect events in multimedia streams.
			
			
			To detect collective anomalies,~\cite{caudell1993adaptive} adopts the \textit{ART (Adoptive Resonance Theory)} neural networks to detect time-series anomalies. \textit{Box Modeling} is proposed in ~\cite{chan2005modeling}. \textit{Longest Common Subsequence} was leveraged in~\cite{budalakoti2006anomaly} as similarity metric for symbolic sequence. Markovian modeling techniques are also popular in this domain\cite{ye2000markov,warrender1999detecting,pavlov2003sequence}. \cite{yu2015glad} depicts groups in social media as combinations of different ``roles'' and compare groups according to the proportion of each role within each group.
		
			Wang et al. proposed a technique, \textit{Multinomial Goodness-of-Fit} (MGoF), to analyze likelihood ratio of distributions via Kullback-Leibler divergence, and is fundamentally a hypothesis test on distributions~\cite{wang2011statistical}.
			MGoF divides the observed data sequence into several windows. It quantifies data in each window into a histogram and check these estimated distributions against several hypothesis. If the target distribution rejects all provided hypothesis, it is considered an anomaly and preserved as a new candidate of null hypothesis. If the target distribution failed to reject some hypothesis, then it is considered a supporting evidence of the one that yields most similarity. Furthermore, if the number of supporting evidence is larger than a threshold $c_{th}$, it is classified as non-anomaly.
			MGoF is the best competitor out of the similar techniques, and we use it as our baseline against our approach.
	
		\ShrinkedSubsection{Real World Problem: Click Farming Detection}\label{sec:related-realworld}
			Click farming is the behaviour that online sellers use a large number of customer accounts to create fake transaction records and give high remarks on products.
			There are two types of click farming behaviours: centralized and equalized.
			Equalized click farming refers to scenarios where behaviours are well organised while centralized one does not.
			Current detection techniques for click farming mainly focus on user behaviours. Those techniques require platforms to keep records on user features. However, the detection can be easily bypassed by trained workers and some well programmed applications.
			
			Although it is hard to classify users as honest or malicious, we can still find clues from the sellers' aspect.
			No matter how much alike between honest users and malicious workers, the fake transaction records will always cause a bias or distortion of the original transaction distribution.
			Thus, if we can measure the similarity between different transaction distributions, there is still a chance for us to detect dishonest sellers.

	\ShrinkedSection{Statistical Detection}\label{sec:algorithm-details}
		\ShrinkedSubsection{Statistical Divergence Detection with Reference (SDD-R)}\label{sec:alg-opt-reference}
			Distortion or drift of certain distribution can be quantified by statistical divergence.
			It provides a distance between two or more distributions. In a set of data collections, we can only draw a complete graph where nodes denote data collections and edges refer to the symmetric divergence between two compared nodes. From the graph we can find some points that have apparently larger distances with most of other points and return them as anomalies. This may work if anomalous nodes do not compose a large proportion. However the procedure will be too complicated to work out with large amounts of data. If it is assured that data collections form only one cluster, some optimizations can be applied to reduce complexity.
			
			Alternatively we can provide a frame of reference that generates absolute coordinates rather than the relative ones. This optimization is feasible if data collections form one single cluster in distribution space. This is true in most reality scenarios given that distribution is adopted to depict a macro property which comes out as one universal conclusion. In other words , if multiple distributions are used to describe subgroups of entire sample space, then a conclusive one can be obtained by averaging all these sub-distributions. Therefore, we can use an estimated cluster center as reference and test distances between the reference and each other data collections(Algorithm~\ref{alg:sdd-r}), yielding absolute distances.
	
			\begin{algorithm}[!t]
				\caption{SDD-R}
				\label{alg:sdd-r}
				\begin{algorithmic}[1]
					\Require Data Collections $\mathbb{D} = \{D_1, \dots, D_n\}$
					\Require Divergence metric $div$
					\Ensure Anomalous Data Collections
					\For {$i \gets 1$ to $n$}
					\State $P_i \gets$ the distribution of $D_i$
					\EndFor
					\State $P_R \gets \frac{1}{n}\sum_{i = 1}^n P_i$
					\For {$i \gets 1$ to $n$}
					\State $d_i \gets div(P_i||P_R)$
					\EndFor
					\State $\mathcal{N}(\mu, \sigma) \gets$ Gaussian distribution estimated by $d_i$
					\State \Return $\{D_i | \frac{d_i - \mu}{\sigma} > 3 \}$
				\end{algorithmic}
			\end{algorithm}
	
			Distribution of all divergences against the reference can be approximated as a Gaussian distribution though the true one may differ a little more from the standard Gaussian than the expected estimation error. That is due to the unknown randomness within real world data. Few assumptions can be applied in real world data sets, no mention that data volume is sometimes relatively low. This topic is out of the domain discussed in this paper and we here only introduce the technique instead of the specific distribution model. 
			Certainly, if stronger assumptions can be included to provide a more precise model, this component in the framework can be replaced to give better results.
			For the simplicity of our proposal, we deem the distributions of divergences to be Gaussian.
			
\begin{figure}[!t]
				\centering
				\includegraphics[width=\linewidth]{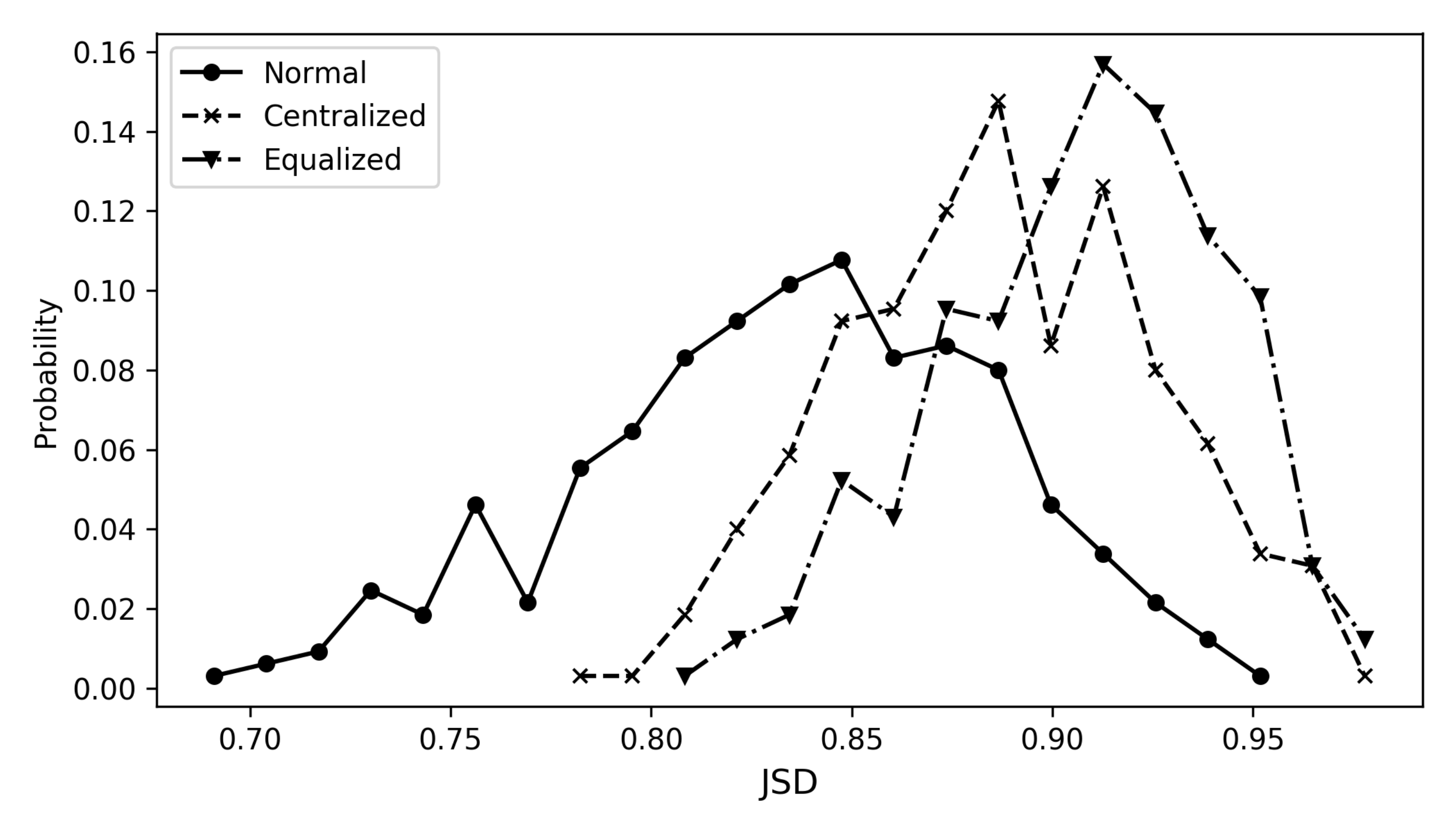}
				\caption{This figure shows distribution of JSD values(on 2nd level histograms) of normal and two types of click farming data. Divergences were calculated according to a reference averaged among all correct distributions.}
				\label{fig:raw-overview}
			\end{figure}

			By this approach, time complexity can be reduced from quadratic to linear. Fig.~\ref{fig:raw-overview} in Section~\ref{sec:exp-raw} demonstrates the result of the above process. Solid line refer to the distances calculated from normal data collections, dashed and dash-dotted ones are from click-farmed data collections. Clearly, distances of normal data collections assembles together around a small value while anomalous ones lay around a larger distance value.
	
		\ShrinkedSubsection{Optimization: Statistical Divergence Detection with Evidence(SDD-E)}
			It is possible to further optimize SDD-R if we can provide this algorithm with evidence(Algorithm~\ref{alg:sdd-e}).
	
			\begin{algorithm}[!t]
				\caption{SDD-E}
				\label{alg:sdd-e}
				\begin{algorithmic}[1]
					\Require Evidence set with normal data collections $\mathbb{E}_N = \{N_1, \dots, N_n\}$
					\Require Evidence set with anomalous data collections $\mathbb{E}_A = \{A_1, \dots, A_m\}$
					\Require Divergence metric $div$
					\Require Estimated anomalous probability $\alpha$
					\Require New data collection $\mathbb{D} = \{D_1, \dots, D_l\}$
					\Ensure Anomalous data collections in $\mathbb{D}$
					\For{each $N_i\in \mathbb{E}_N, A_j\in \mathbb{E}_A, D_k\in \mathbb{D}$}
					\State $P_{N_i} \gets$ distribution of $N_i$\label{line:hist-1}
					\State $P_{A_j} \gets$ distribution of $A_j$\label{line:hist-2}
					\State $P_k \gets$ distribution of $D_k$
					\EndFor
					\State $P_R \gets \frac{1}{n}\sum_{i=1}^{n}P_{N_i}$
					\For{each $N_i\in \mathbb{E}_N, A_j\in \mathbb{E}_A$}
					\State $d_{N_i} \gets div(P_{N_i}||P_R)$
					\State $d_{A_j} \gets div(P_{A_j}||P_R)$
					\EndFor
					\State $\mathcal{N}_N(\mu_N, \sigma_N) \gets$ normal distribution estimated from $\{d_{N_1}, \dots, d_{N_n}\}$
					\State $\mathcal{N}_A(\mu_A, \sigma_A) \gets$  normal distribution estimated from $\{d_{A_1}, \dots, d_{A_m}\}$
					\State $T \gets$ proper threshold derived from $\mathcal{N}_N$, $\mathcal{N}_A$ and $\alpha$
					\State \Return $\{D_i | div(P_i||P_R) > T\}$
				\end{algorithmic}
			\end{algorithm}
	
			Evidences enables the algorithm to not only refine estimation of real distribution but also build knowledge of anomalous collections, which is similar to the parameter estimation within a certain sample set.
			
			According to the property of statistical divergence, we can infer that the true distribution of divergences calculated from normal data collections are close to but not exactly a Gaussian distribution $\mathcal{N}(\mu, \sigma)$ since for each point, there are both definite upper and lower bounds instead of infinities. Therefore, $\mu$ should be slightly larger than zero($\mu = 0 \iff P_i = P_j, \forall P_i, P_j \in \mathbb{E}_N$, for real world data sets, this is highly unlikely). Time complexity for this algorithm is still linear but with a larger coefficient.
			
			For certain divergence, it is possible to compare similarity from one distribution against multiple others, such as Jensen-Shannon Divergence. Although it reduces time complexity, it sacrifices unaffordable accuracy because divergence among multiple distribution dilutes differences. Take JSD as an example, suppose $P(1) = P(2) = P(3) = \frac{1}{3}$ and $Q(1) = \frac{1}{6}, Q(2) = \frac{1}{3}, Q(3) = \frac{1}{2}$, then $JSD(P||Q) \approx 0.033$ and $JSD(P, P, P, Q) \approx 0.024$.
	
			This algorithm can be slightly modified to deal with concept drift(for example, trading trend changes over time for online shops as they are often in the process of expanding or dwindling) by turning the two evidence sets as sliding windows and adopting certain update strategies such as \textit{Least Recently Used}(LRU). Time complexity for this optimization is $O(n\cdot(|\mathbb{E}_N|+|\mathbb{E}_A|)\cdot T_D)$, where $T_D$ denotes time complexity of divergence calculation.
			
		\ShrinkedSubsection{Threshold}\label{sec:alg-threshold}
			\begin{figure}[t]
				\centering
				\includegraphics[width=\linewidth]{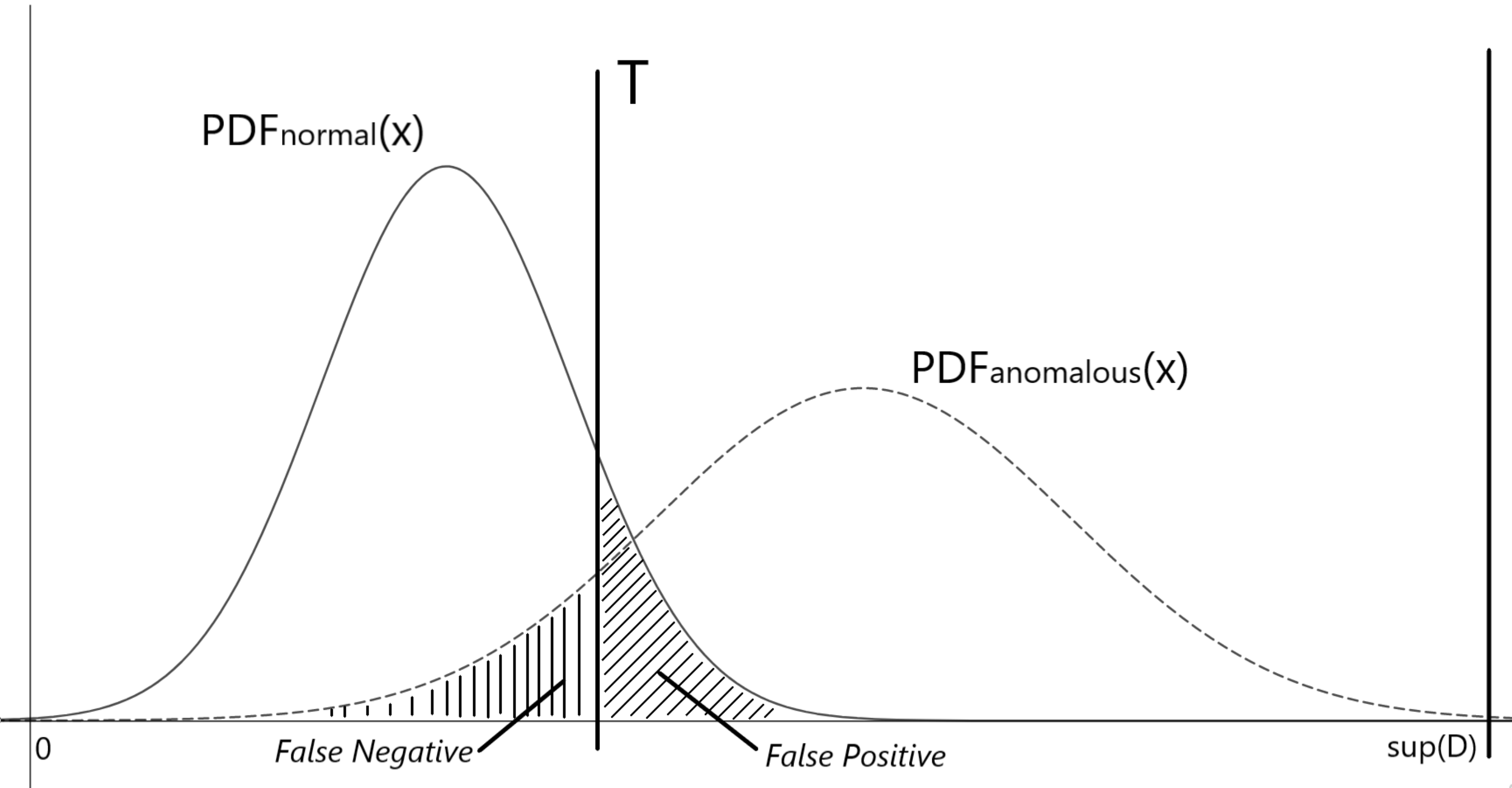}
				\caption{Optimal threshold should minimize the size of shadow under curve.}
				\label{fig:example-threshold}
			\end{figure}
	
			One important factor in algorithm SDD-E is the value of threshold.
			A naive but prevalent approach is to set a fixed value as the threshold(As is shown in Algorithm~\ref{alg:sdd-r}).
			However, a fixed threshold requires specific analysis in the certain scenario, manual observations and tuning of parameters, which involves lots of human labour.
			
			An adaptive threshold is chosen in our technique to minimize total errors(both false negative and false positive).
			However, this is not accurate enough, it implicates an assumption that chances are the same for a new data collection to be either anomalous or not. If we can determine the probability for a new data collection to be anomalous in any segment of data sequence, the equation should be modified as minimizing expected errors, where we use $\alpha$ to denote anomaly probability.
			Suppose: $PDF_{normal}(x) \sim \mathcal{N}(\mu_n, \sigma_n)$ and $PDF_{anomalous}(x) \sim \mathcal{N}(\mu_a, \sigma_a)$, then adaptive threshold can be calculated by E.q.(\ref{equ:linear-weight}).
			
			Moreover, with an estimated anomaly probability, SDD-R can be also optimized by ranking all data collections according to their divergence value and select first $n \cdot \alpha$ ones with highest values as anomalies.
	
			\begin{table*}
				\footnotesize
				\begin{align}\label{equ:linear-weight}
				T & = \mathop{\arg\min}_{T} \alpha\int_{0}^{T}PDF_{a}(x)dx +
				(1-\alpha)\int_{T}^{\sup(D)}PDF_{n}(x)dx\nonumber\\
				& \approx \mathop{\arg\min}_{T}
				\alpha\int_{-\infty}^{T}
				\frac{e^{-\frac{(x - \mu_a)^2}{2\sigma_a^2}}}{\sqrt{2\pi} \sigma_a}dx
				+ 
				(1-\alpha)\int_{T}^{+\infty}
				\frac{e^{-\frac{(x - \mu_n)^2}{2\sigma_n^2}}}{\sqrt{2\pi} \sigma_n}dx\nonumber\\
				& = \begin{cases}
				\displaystyle
				\frac{1}{\sigma_a^2 - \sigma_n^2}\left[(\sigma_a^2\mu_n - \sigma_n^2\mu_a) \pm \sigma_a\sigma_n\sqrt{(\mu_a - \mu_n)^2 + 2(\sigma_a^2 - \sigma_n^2)ln\frac{(1 - \alpha)\sigma_a}{\alpha\sigma_n}}\right], & \sigma_a \ne \sigma_n\\
				\displaystyle
				\frac{\mu_n + \mu_a}{2} + \frac{k^2ln\frac{1 - \alpha}{\alpha}}{\mu_a - \mu_n}, & \sigma_a = \sigma_n = k
				\end{cases}
				\end{align}
				
				Note: when $\sigma_a \ne \sigma_n$, keep the root s.t. $\displaystyle \frac{\alpha (T - \mu_a)}{\sigma_a^3}e^{-\frac{(T - \mu_a)^2}{2\sigma_a^2}} < \frac{(1 - \alpha) (T - \mu_n)}{\sigma_n^3}e^{-\frac{(T - \mu_n)^2}{2\sigma_n^2}}$
			\end{table*}
	
	\ShrinkedSection{Evaluation}\label{sec:evaluation}
		Our algorithm was implemented and interpreted in Python 3.5.2. All experiments were tested on Ubuntu 16.04. In the following experiments, we figured out properties of real world data and performance of our technique against anomalous data collections. We also made a comparison among variations of SDD algorithms and MGoF.
		\footnote{All resources and more detailed experiment results can be viewed in supplemental material.}
	
		\ShrinkedSubsection{Methodology}\label{sec:exp-methodology}
			We adopted Koubei sellers' transaction records\footnote{https://tianchi.aliyun.com/competition/information.htm?raceId=231591} in experiments. It was provided by Alibaba Tian Chi big data competition where all records were collected from real world business scenarios. It contains transaction records of 2000 sellers from 2015-07-01 to 2016-10-31.
			Two types of click-farmed data was generated according to patterns described in section~\ref{sec:related-realworld}.
			In our experiments, we use $\nu$ to denote the magnitude coefficient of click farming. Hence $|D_{anomalous}| = (1 + \nu)|D_{normal}|$. In the following experiments without extra illustration, we adopted $\nu = 1$.
			
			One defect of this data set is that the detailed time stamp is aligned at each hour of the day due to desensitization. We constructed an enhanced data set by assigning every time stamp a random value for minutes and seconds. Therefore, the enhanced data set should be closer to the reality.
	
			Divergence metric adopted in each SDD algorithms was Jensen-Shannon divergence if no specific notation is made. However, MGoF used only Kullback-Leibler divergence due to its special mechanism. We use a ``+'' to denote algorithms optimized by a given $\alpha$.
	
		\ShrinkedSubsection{Experiments on Koubei Data Set}\label{sec:exp-raw}
			We first tested our algorithms to see whether and why the algorithm works. Anomalies were random selected days replaced by corresponding click farmed version. To play the role of purchasing platform, we investigated two levels of transaction distribution. The first level is to simply draw a histogram aligned to time spans. The second level is to draw a histogram on the sub-volumes in each time span(i.e. a histogram on frequencies in the first level histogram).
			To test SDD-E, we randomly selected 30 correct days and 10 click farmed days as normal and anomalous evidence respectively.
			Here $\alpha = 0.2$. The results are shown in Table~\ref{tab:result-koubei-raw} and \ref{tab:result-koubei-enhanced}.
	
			\begin{sidewaystable}
				\centering
				\caption{Performance on Raw Data}
				\label{tab:result-koubei-raw}
				\footnotesize
				\begin{tabular}{|c|c|c|c|c|c|c|c|c|c|c|c|c|c|c|c|c|}
					\hline
					\multirow{3}{*}{\textbf{}} & \multicolumn{8}{c|}{\textbf{Centralized}} & \multicolumn{8}{c|}{\textbf{Equalized}} \\ \cline{2-17} 
					& \multicolumn{4}{c|}{\textbf{1st Level}} & \multicolumn{4}{c|}{\textbf{2nd Level}} & \multicolumn{4}{c|}{\textbf{1st Level}} & \multicolumn{4}{c|}{\textbf{2nd Level}} \\ \cline{2-17} 
					& \textbf{Pre(\%)} & \textbf{Rec(\%)} & \textbf{F1(\%)} & \textbf{T(ms)} & \textbf{Pre(\%)} & \textbf{Rec(\%)} & \textbf{F1(\%)} & \textbf{T(ms)} & \textbf{Pre(\%)} & \textbf{Rec(\%)} & \textbf{F1(\%)} & \textbf{T(ms)} & \textbf{Pre(\%)} & \textbf{Rec(\%)} & \textbf{F1(\%)} & \textbf{T(ms)} \\ \hline
					\textbf{SDD-R} & 89.51 & 48.75 & 63.12 & 266.77 & 21.97 & \textbf{99.38} & 35.98 & 11.12 & 6.67 & 0.63 & 1.14 & 249.15 & 21.22 & 72.50 & 32.83 & 7.81 \\ \hline
					\textbf{SDD-R+} & 91.25 & 91.25 & \textbf{91.25} & \textbf{265.96} & \textbf{61.88} & 61.88 & 61.88 & 10.08 & 9.38 & 9.38 & 9.38 & \textbf{247.31} & \textbf{44.38} & 44.38 & 44.38 & 6.92 \\ \hline
					\textbf{SDD-E Static} & \textbf{92.46} & 68.75 & 78.86 & 292.50 & 36.55 & 98.13 & 53.26 & 5.75 & 6.67 & 0.63 & 1.14 & 271.45 & 36.07 & 86.25 & 50.86 & 5.64 \\ \hline
					\textbf{SDD-E Static+} & 85.02 & 32.50 & 47.02 & 293.77 & 46.24 & 91.88 & 61.52 & 5.95 & 10.00 & 0.63 & 1.18 & 272.71 & 43.60 & 76.25 & \textbf{55.48} & 5.68 \\ \hline
					\textbf{SDD-E Dynamic} & 49.11 & \textbf{99.38} & 65.73 & 699.97 & 23.01 & \textbf{99.38} & 37.37 & 245.65 & 10.36 & \textbf{18.13} & \textbf{13.18} & 681.09 & 22.09 & \textbf{93.13} & 35.71 & 242.85 \\ \hline
					\textbf{SDD-E Dynamic+} & 73.21 & 98.75 & 84.09 & 701.06 & 48.02 & 96.25 & \textbf{64.07} & 255.43 & 8.15 & 6.88 & 7.46 & 681.89 & 40.79 & 78.13 & 53.59 & 253.03 \\ \hline
					\textbf{MGoF} & 14.08 & 21.88 & 17.13 & 292.14 & 13.01 & 4.38 & 6.55 & \textbf{3.64} & \textbf{12.50} & 3.13 & 5.00 & 250.42 & 12.50 & 3.13 & 5.00 & \textbf{3.71} \\ \hline
				\end{tabular}
				
				\vskip\baselineskip
				\caption{Performance on Enhanced Data}
				\label{tab:result-koubei-enhanced}
				\footnotesize
				\begin{tabular}{|c|c|c|c|c|c|c|c|c|c|c|c|c|c|c|c|c|}
					\hline
					\multirow{3}{*}{\textbf{}} & \multicolumn{8}{c|}{\textbf{Centralized}} & \multicolumn{8}{c|}{\textbf{Equalized}} \\ \cline{2-17} 
					& \multicolumn{4}{c|}{\textbf{1st Level}} & \multicolumn{4}{c|}{\textbf{2nd Level}} & \multicolumn{4}{c|}{\textbf{1st Level}} & \multicolumn{4}{c|}{\textbf{2nd Level}} \\ \cline{2-17} 
					& \textbf{Pre(\%)} & \textbf{Rec(\%)} & \textbf{F1(\%)} & \textbf{T(ms)} & \textbf{Pre(\%)} & \textbf{Rec(\%)} & \textbf{F1(\%)} & \textbf{T(ms)} & \textbf{Pre(\%)} & \textbf{Rec(\%)} & \textbf{F1(\%)} & \textbf{T(ms)} & \textbf{Pre(\%)} & \textbf{Rec(\%)} & \textbf{F1(\%)} & \textbf{T(ms)} \\ \hline
					\textbf{SDD-R} & 81.54 & 41.88 & 55.33 & 238.61 & 17.49 & 92.50 & 29.42 & 13.86 & 6.67 & 0.63 & 1.14 & 239.44 & 21.15 & 71.88 & 32.69 & 12.08 \\ \hline
					\textbf{SDD-R+} & \textbf{91.25} & 91.25 & \textbf{91.25} & \textbf{236.94} & 36.88 & 36.88 & 36.88 & 12.98 & 8.13 & 8.13 & 8.13 & \textbf{236.47} & 38.75 & 38.75 & 38.75 & 11.03 \\ \hline
					\textbf{SDD-E Static} & 88.31 & 67.50 & 76.52 & 257.60 & 31.99 & 92.50 & 47.54 & 9.74 & 6.67 & 0.63 & 1.14 & 257.65 & 34.49 & 91.25 & 50.06 & 9.37 \\ \hline
					\textbf{SDD-E Static+} & 69.67 & 13.13 & 22.09 & 259.30 & \textbf{42.12} & 73.75 & \textbf{53.62} & 9.59 & 10.00 & 0.63 & 1.18 & 258.45 & \textbf{44.17} & 82.50 & \textbf{57.54} & 9.32 \\ \hline
					\textbf{SDD-E Dynamic} & 42.93 & \textbf{99.38} & 59.96 & 1106.25 & 20.12 & \textbf{95.63} & 33.25 & 277.70 & 10.57 & \textbf{17.50} & \textbf{13.18} & 1108.93 & 20.30 & \textbf{94.38} & 33.42 & 271.74 \\ \hline
					\textbf{SDD-E Dynamic+} & 69.18 & 98.13 & 81.15 & 1110.81 & 33.25 & 87.50 & 48.19 & 292.60 & 7.06 & 3.75 & 4.90 & 1118.00 & 37.21 & 80.63 & 50.92 & 283.50 \\ \hline
					\textbf{MGoF} & 13.97 & 17.50 & 15.54 & 294.08 & 14.66 & 8.13 & 10.45 & \textbf{8.01} & \textbf{12.50} & 3.13 & 5.00 & 250.10 & 18.42 & 8.75 & 11.86 & \textbf{7.55} \\ \hline
				\end{tabular}
			\end{sidewaystable}
	
			When classifying toward 1st level histograms, centralized click farming behaviours can be easily discovered. It is because normal collections share a similar distribution while centralized click-farmed ones abruptly violated the original shape.
			When it came to 2nd level histograms, equalized click farming were also effectively discovered. It can be clearly seen in Fig.~\ref{fig:raw-overview} that distribution of divergence of both click farming types shows an obvious deviation from the normal one.
			
			The result showed that our technique outperformed MGoF in every real world case.
			SDD-E provided best performance, yet it consumed the most computing power. Comparison between SDD-R and MGoF revealed improvement of reference as well as the importance of threshold under this technique. It is also clear that dynamic SDD-E is capable of tracing the gradual shift of environment. MGoF turned out to be the worst since it always mark several false positive when $c_{th}$ had not been met and much more false negatives when similar errors occurred too many. 
			
			Parameter $\alpha$ improved total accuracy of dynamic SDD-E algorithm by 10-20\% as was supposed. It also increased its F1 by more than 20\%. $\alpha$ made a great difference in SDD-R as well, which illustrated that divergence sorted almost all collections in correct order according to the averaged reference. However, static SDD-E did not show the same improvement. Since environment drift took greater influence in the result. In comparison with $\alpha$, adaptive threshold given by evidence sets did not bring the most improvement. But this threshold can be applied together with other optimizations such as slide windows.
	
		\ShrinkedSubsection{Test against Anomaly Proportion and Magnitude}\label{sec:exp-synthetic}
			In this experiment, we tested algorithm performance under various anomaly proportion and magnitude. $\alpha$ ranged from 0.1 to 0.9 when $\nu = 1$ and $\nu \in [0.1, 0.9]$ when $\alpha = 0.1$, other settings remains the same.
	
			\begin{figure}[!t]
				\includegraphics[width=\linewidth]{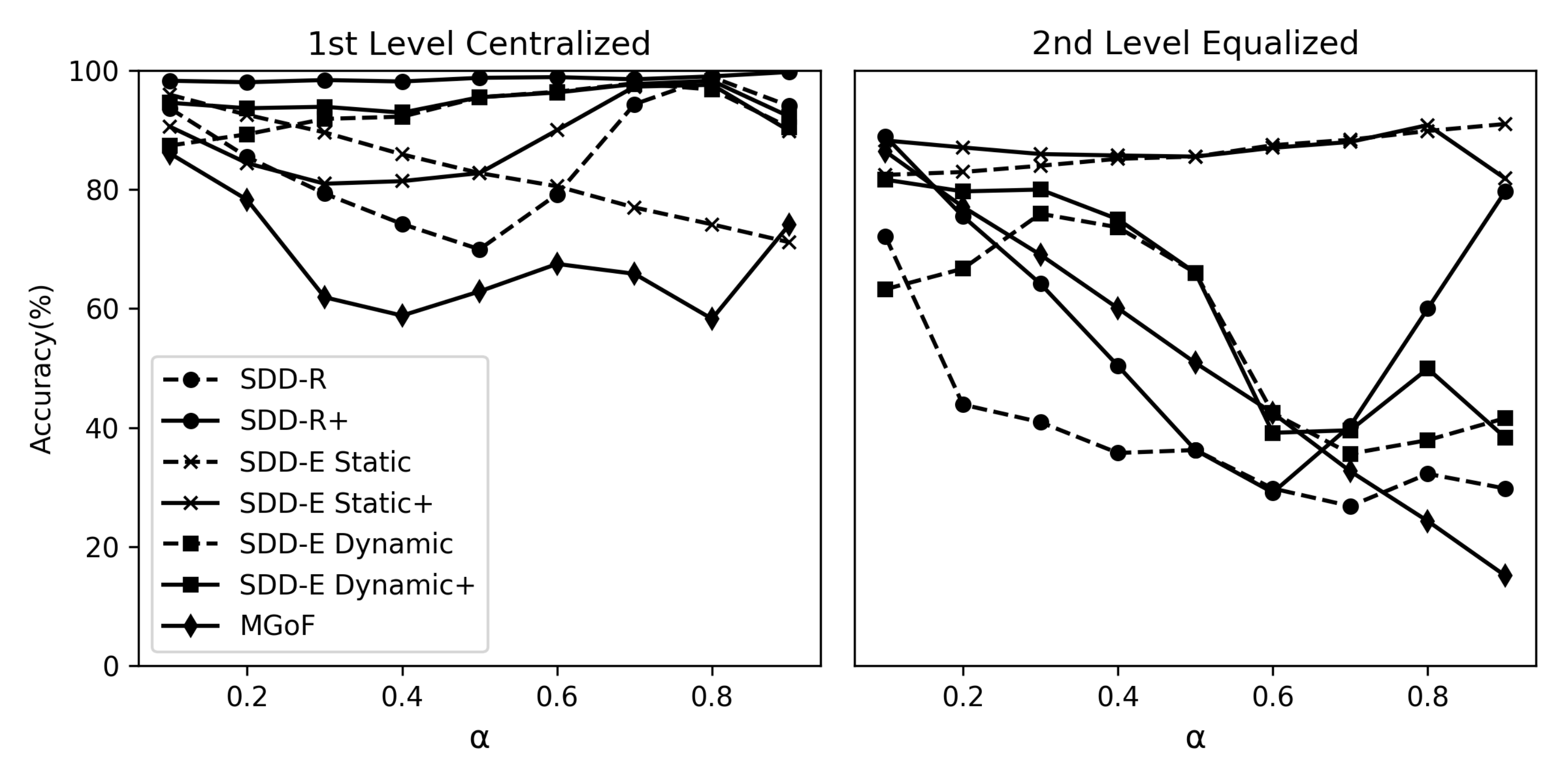}
				
				\includegraphics[width=\linewidth]{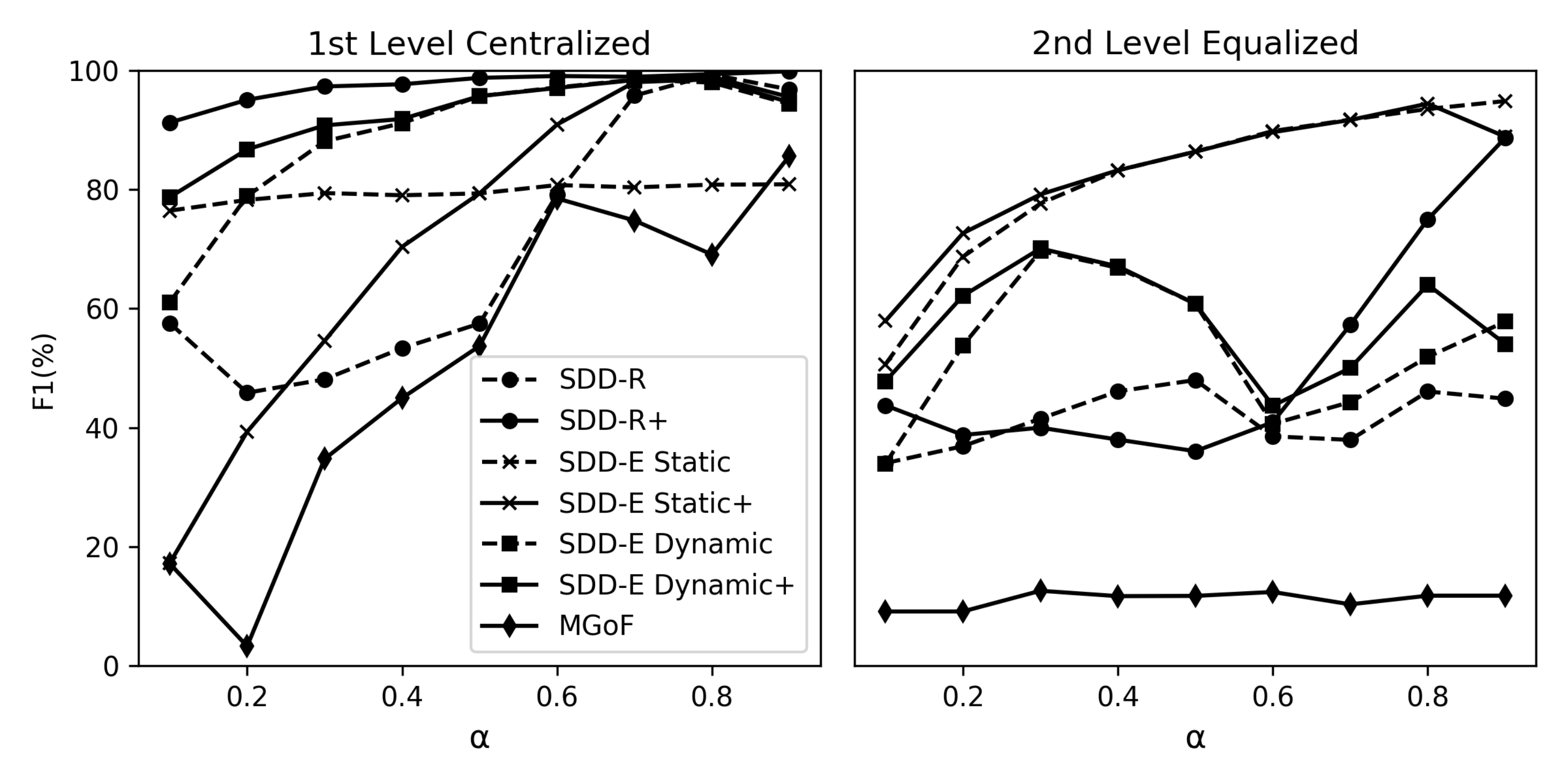}
				\caption{Accuracy and F1 on Different Anomaly Probabilities}
				\label{fig:anomaly-probability}
			\end{figure}
	
			Fig.~\ref{fig:anomaly-probability} shows that our technique outperformed MGoF and was relatively stable when dealing with all proportions of 1st level centralized anomalies. SDD-E performed even better since it maintains knowledge of both normal and anomalous distributions and calculates the threshold according to the best expectation. However, it relies on the accuracy of distribution estimation. When it came to 2nd level distributions, histograms became much coarser since data available was highly limited and thus its performance suffered dramatically.
			For the classifiers of MGoF, they  compromised to a high error rate. Because more anomalies gathered together and the algorithm recognized them as clusters of normal data.
	
			\begin{figure}[!t]
				\includegraphics[width=\linewidth]{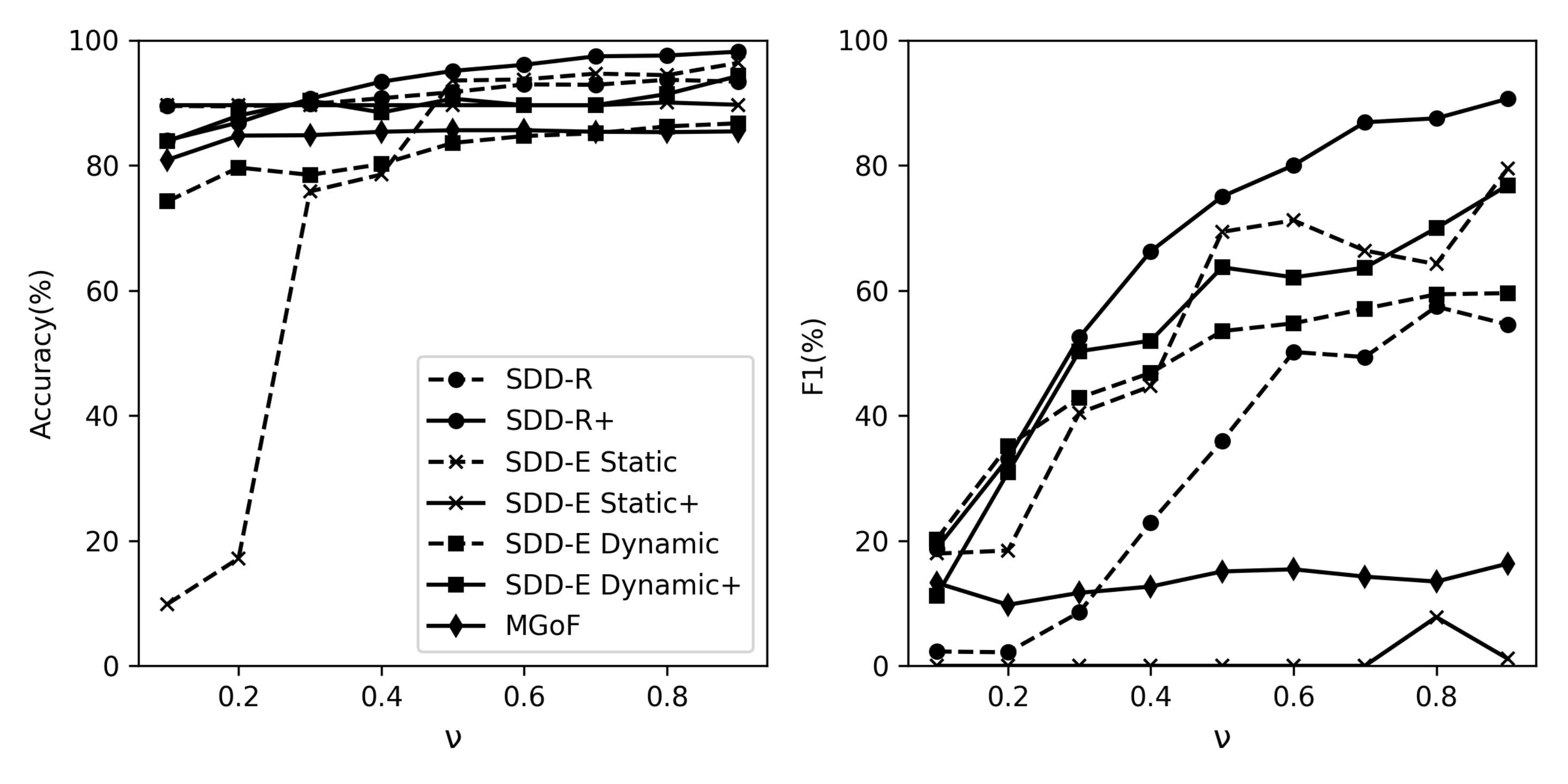}
				\caption{Accuracy and F1 on Different Anomaly Magnitudes}
				\label{fig:anomaly-magnitude}
			\end{figure}
	
			From Fig.~\ref{fig:anomaly-magnitude} we can conclude that our algorithms are still the best, given that they are most sensitive toward tiny anomalous variations. However, static SDD-E did not rise until $\nu > 1$, this is because it suffered from fluctuation on the trade environment at the mean time. MGoF is not sensitive toward minor anomalies either.
			For a relatively small magnitude of click farming, the classifiers of MGoF quickly degrade to be trivial. The rigid threshold could not automatically rise up and was thus far from optimal. 
	
		\ShrinkedSubsection{Discussion}
			MGoF's learning procedure of anomalous probability hypothesis is inefficient. To maintain a comprehensive knowledge of anomalies, MGoF has to reserve a single hypothesis entry for every type of them. But in reality, it is always the case that we face the heterogeneity of outliers. In the Koubei data set, there can be tens of anomalous distributions caused solely by centralized click farming. It takes a long time to discover every possible type of anomaly. Besides, if there happens to be more than $c_{th}$ anomalous distributions of the same type, later discovered collections will no longer declared to be anomalous any more.
			
			However, in SDD-R and SDD-E, that is not a problem since it can map and gather all anomalies together and draw a universal boundary between them and all normal collections. These techniques are suitable to all typical divergence metrics and consume little computation power(except dynamic SDD-E). The only drawback is that they require comprehensive estimation of target distributions. Although other parameters need estimation as well, they are naturally addressable under big data circumstances.
	
	\ShrinkedSection{Conclusion}\label{sec:conclusion}
		This paper proposes a series of collective anomaly detection techniques, which helps detect data manipulations in modern data pipelines and data centres. Different from existing algorithms designed for collective anomalies, our approach employs statistical distance as the similarity measurement. We explored several technical points involved in the design of the algorithm and performed a thorough experiment to test its efficiency. The comparison experiment also illustrated the advantages of our technique. It can be concluded that our technique can efficiently discover anomalies within the data collections and the classifier is sensitive enough toward real world data manipulations. 
		
	\subsection*{Acknowledgement}
		This is work is supported by the National Science Foundation of China with Grant 61872232 and Shanghai Pujiang Program with 19pj1433100.

	\bibliographystyle{IEEEbib}
	\bibliography{CameraReadyVersion}

\end{document}